# Finite Size Effects in Fluid Interfaces*

M. Caselle[a], F. Gliozzi[a], P. Provero[b] and S. Vinti[a, c]

[a]Dip. di Fisica Teorica dell'Università di Torino and I.N.F.N., Via P. Giuria 1, I-10125 Torino, Italy

[b]Theoretical Physics Institute, University of Minnesota, Minneapolis, MN 55455, USA

[c]Centro Brasileiro de Pesquisas Fisicas, Rua Dr. Xavier Sigaud 150, 22290 Rio de Janeiro, Brazil

It is shown that finite size effects in the free energy of a rough interface of the 3D Ising and three–state Potts models are well described by the capillary wave model at *two–loop* order. The agreement between theoretical predictions and Monte Carlo simulations strongly supports the idea of the universality of this description of order–order interfaces in 3D statistical systems above the roughening temperature.

## 1. Introduction

Many 3D physical systems show the presence of interfaces separating coexisting phases at thermal equilibrium. It turns out that these interfaces are often dominated by long wavelength fluctuations (*i.e.* they behave as *fluid* interfaces). These soft modes play an essential role in the description of *finite size effects* (FSE) in the fluid interface's free energy (see *e.g.* [1]).

3D spin systems offer a simple context where these effects appear and can be studied, by using numerical simulations to check theoretical predictions. It is in fact well known that at low temperature, on finite volumes, they show domain walls separating coexisting phases, which behave as fluid interfaces between the critical and the roughening temperature. Besides, the models we study are interesting because they are related by duality to 3D $Z_N$ gauge theories, and to high temperature 4D gauge theories through dimensional reduction.

While below the roughening temperature the interfaces are almost rigid and a theoretical microscopical approach can be taken (see *e.g.* [2] and references therein), above it one is forced to assume an effective model describing the collective degrees of freedom of the rough interfaces.

We follow the *capillary wave model* (CWM) [3], in its simplest formulation, assuming an effective hamiltonian proportional to the area $A[x]$ of the surface

$$A\,[x] = \int_0^R dr \int_0^T dt\,\sqrt{1 + \left(\frac{\partial x}{\partial r}\right)^2 + \left(\frac{\partial x}{\partial t}\right)^2} \quad (1)$$

where the single–valued function $x(r,t)$ describes the displacement from the equilibrium position of the interface and $\sigma$ is the reduced (order–order) interface tension. To compare the predictions of the CWM with numerical results obtained from Monte Carlo (MC) simulations, one can choose 3D lattices of $R \times T \times L$ sites, with $L \gg R, T$ and periodic boundary conditions in each direction. This particular choice of the lattice shape allows one to consider only interfaces orthogonal to the elongated direction $L$, the probability of having interfaces orthogonal to the other directions being negligible. The 2D field $x(r,t)$ is therefore defined on the rectangle $(r,t) \in [0,R] \times [0,T]$ with opposite edges identified, *i.e.* on a torus.

## 2. The *two–loop* CWM approximation

Rather strong FSE, depending on the shape of the lattice, already appear at the *one–loop* (gaussian) approximation to the CWM [4]. The partition function $Z_{cw} = \int [Dx]\, e^{-\sigma A[x]}$ can be in fact expressed as an expansion in powers of the adimensional parameter $\sigma_a^{-1}$, proportional to the minimal area of the surface ($\sigma_a \equiv \sigma RT$),

$$Z_{cw}(z,\sigma_a) = \delta e^{-\sigma_a} \cdot Z_{(1l)}(z) \cdot Z_{(2l)}(z,\sigma_a) \ldots \quad (2)$$

*Presented by S. Vinti

where $\delta$ is an unknown constant and $z = R/T$.

The one–loop contribution $Z_{(1l)}$, obtained retaining only the quadratic term in the expansion of $A[x]$, depends only on $z$, namely on the asymmetry of the transverse sizes of the elongated lattice. $Z_{(1l)}$ is nothing but the (exact) partition function of a 2D conformal invariant free boson on a torus of modular parameter $\tau = iz$: $Z_{(1l)}(z) = \frac{1}{\sqrt{z}} |\eta(iz)/\eta(i)|^{-2}$, where $\eta$ is the Dedekind function $\eta(\tau) = q^{1/24} \prod_{n=1}^{\infty} (1 - q^n)$, $q = e^{2\pi i \tau}$ (see *e.g.* [4] and references therein). The one–loop CWM prediction has been tested, on asymmetric lattices ($R \neq T$), in the scaling region of the 3D Ising model and a quite remarkable agreement with the MC data has been found [4].

Taking into account higher order corrections to the gaussian model a more stringent test on the CWM can be provided: in fact, while many different effective hamiltonians reduce to the gaussian form at one–loop level [5], they differ in the form of two– and higher–loop corrections.

The *two–loop* term $Z_{(2l)}$ of Eq.(2) can be calculated perturbatively, expanding (1) at the next–to–leading order in $\sigma_a$, and is given by [6]

$$Z_{(2l)}(z, \sigma_a) = 1 + \frac{1}{2\sigma_a} f(z) + O\left(\sigma_a^{-2}\right) \quad , \qquad (3)$$

where

$$f(z) \equiv \left\{ \left[\frac{\pi}{6} z e_2(iz)\right]^2 - \frac{\pi}{6} z e_2(iz) + \frac{3}{4} \right\} \qquad (4)$$

and $e_2(\tau) = 1 - 24 \sum_{n=1}^{\infty} \frac{n\ q^n}{1-q^n}$, $q = e^{2\pi i \tau}$, is the second Eisenstein series.

In contrast to what happens at the gaussian level, here the capillary wave contributions do not depend only on the asymmetry parameter $z$ but also on $\sigma_a$. As a consequence, the two–loop finite size behavior of the energy splitting $E$ occurring between vacua on finite volumes

$$E(R,T) \equiv Z_{cw}(z, \sigma_a) \qquad (5)$$

has no longer, the well known *classical* functional form [8] $E_{cl}(R) = \delta e^{-\sigma R^2}$ not even for symmetric ($R = T$) lattices, as can be easily seen by setting $z = 1$ into Eq.(5).

## 3. Numerical results

The comparison between formula (5) and the values of $E$ extracted from MC simulations provides a simple and stringent way to verify the CWM predictions. Notice that no new free parameters are introduced within this approach: Eq.(2) contains the same number of undetermined parameters, namely $\sigma$ and $\delta$, to all orders of approximation.

We report here the main results of the MC simulations we have made for the three–state Potts [7] and the Ising model [9]: in both cases we have followed Ref.[10] to extract the energy splittings $E$. For the Potts model, in order to deal only with order–order interfaces, the simulations have been performed at $\beta = 0.3680$ (in the conventions in which $\beta_c \simeq 0.36708$). The results are plotted in Fig.1: the sizes vary in the range $9 \leq T \leq 20$, $10 \leq R \leq 36$, with $R \geq T$, and the longest lattice size is fixed at $L = 120$.

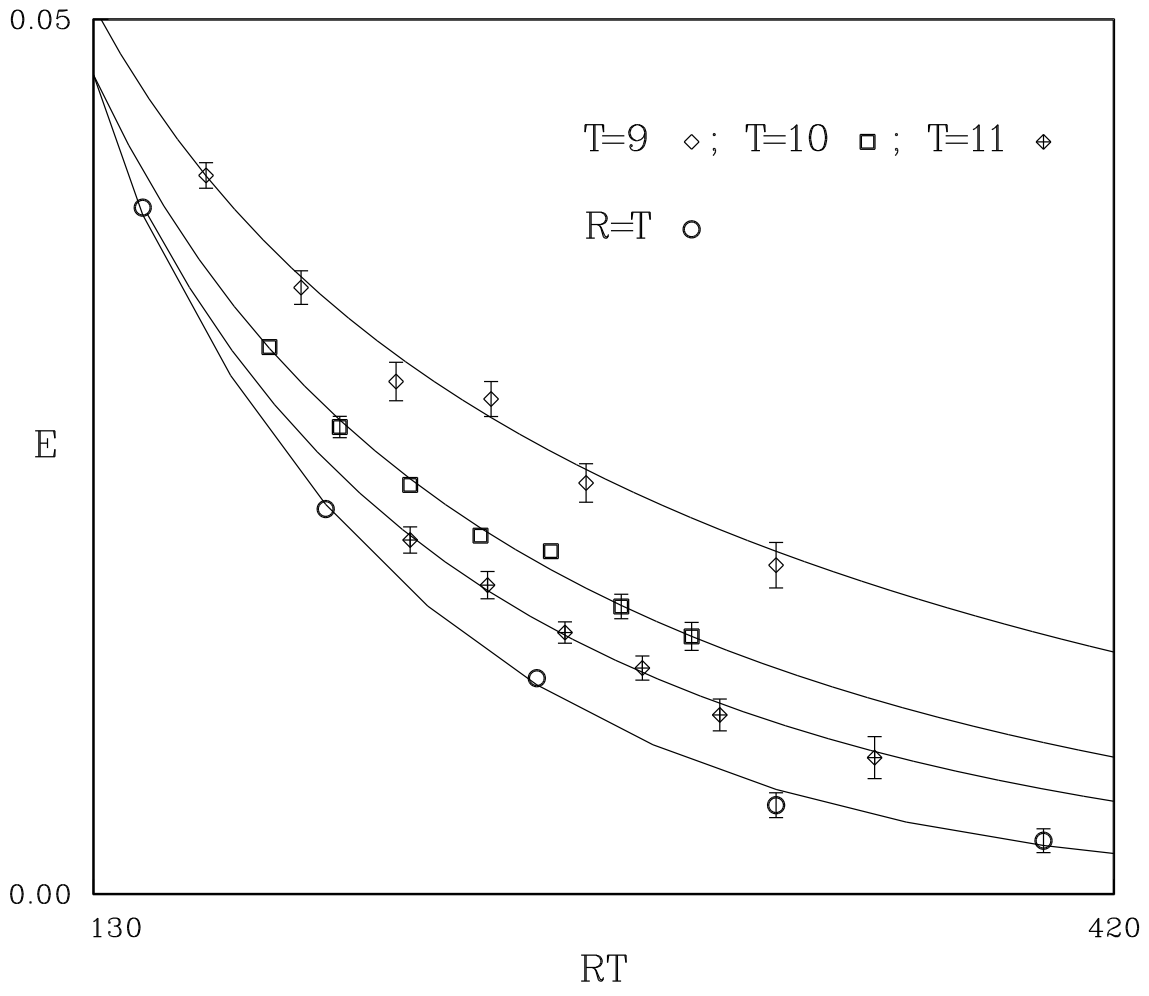

Figure 1. *CWM at two–loop with the MC data for the Potts model at $\beta = 0.3680$. The lines represent the best fit of all data to Eq.(5), $\sigma$ and $\delta$ being given in Tab.1. Error bars not reported are smaller than the plotted symbols.*

For the Ising model, the simulations have been performed at $\beta = 0.2246$ and $\beta = 0.2258$. The results are plotted in Fig.2. It is evident that no single valued function of $\sigma_a$ can describe the data reported in both figures. In particular, within the classical approximation ($Z_{cw} \equiv \delta e^{-\sigma_a}$) of Eq.(5) to fit each sample of data, at fixed $\beta$, one always finds a bad $\chi^2/d.o.f.$. A clear enhancement of the latter can instead be seen if one uses the one–loop approximation ($Z_{cw} \equiv \delta e^{-\sigma_a} Z_{(1l)}(z)$), while good C.L. are reached with the full two–loop form of Eq.(5).

Table 1
Comparison of interface tensions obtained from different theoretical approximations.

| | $\beta$ | approx. | $\sigma$ | $\delta$ | $\chi^2$/d.o.f. |
|---|---|---|---|---|---|
| Potts | 0.3680 | 2-loop | 0.00991(7) | 0.138(2) | 0.73 |
| | | 1-loop | 0.01005(7) | 0.172(2) | 3.60 |
| | | class. | 0.00809(7) | 0.139(2) | 36.3 |
| Ising | 0.2246 | 2-loop | 0.00649(10) | 0.108(2) | 0.58 |
| | | 1-loop | 0.00693(10) | 0.142(3) | 1.65 |
| | | class. | 0.00651(10) | 0.131(3) | 11.1 |
| | 0.2258 | 2-loop | 0.00941(6) | 0.123(2) | 0.44 |
| | | 1-loop | 0.00958(6) | 0.148(2) | 1.48 |
| | | class. | 0.00833(6) | 0.121(2) | 19.2 |

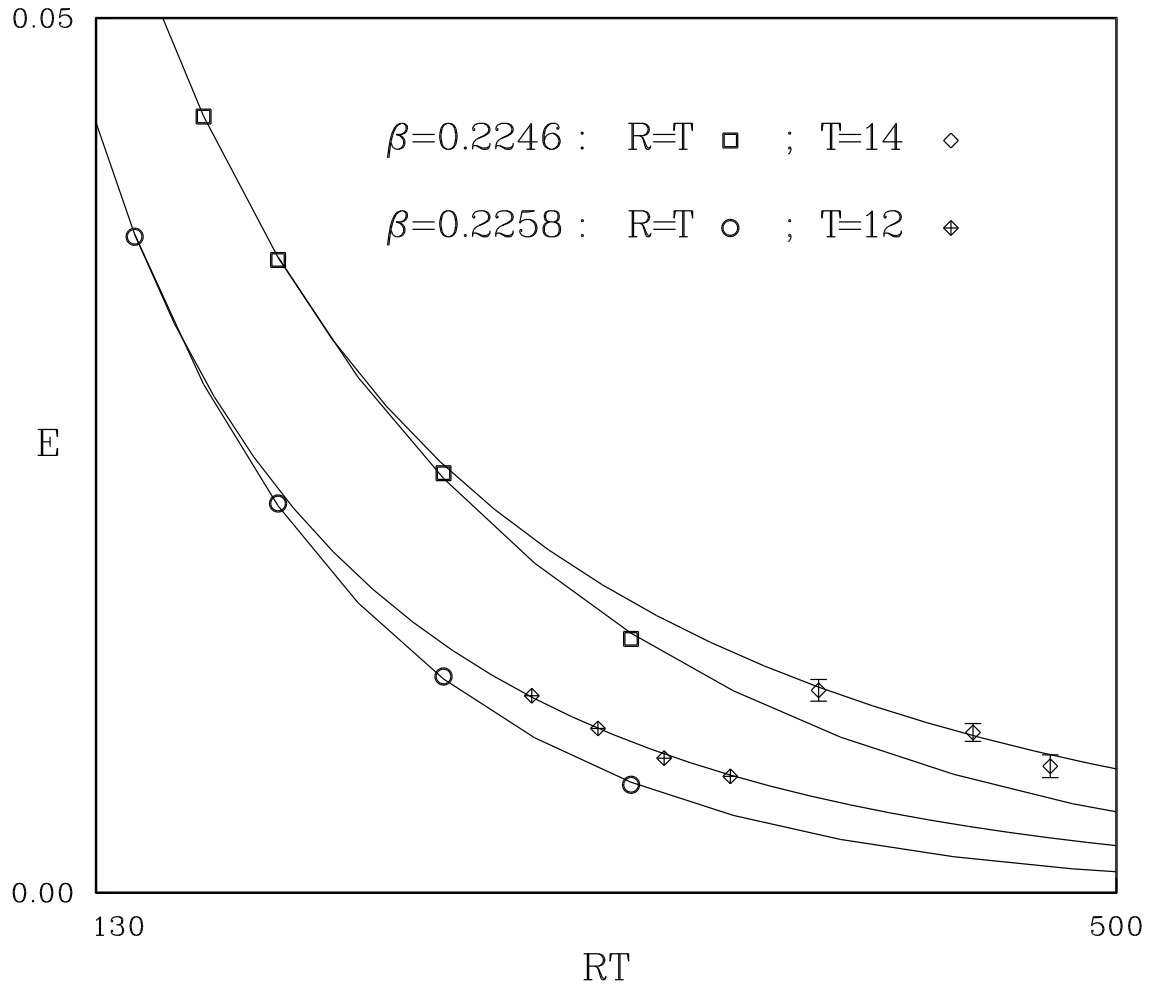

Figure 2. *The same as Fig.1 in the case the Ising model at $\beta = 0.2246$ and $\beta = 0.2258$.*

The results of these fits are given in Tab.1. Thus, the CWM in the two–loop approximation provides an excellent description of order–order interfaces in both the Ising and the three–state Potts models. This result is a strong indication of the universality of this description of interface physics in 3D statistical models.